\documentclass[a4paper,10pt]{article}

\usepackage{graphicx,setspace}
\usepackage{bm}
\usepackage{amsmath,textcomp}

\begin{document}

\title{The role of homophily in the emergence of opinion controversies}

\author{F. Gargiulo$^{1}$, Y. Gandica$^{2}$ \\
\emph{1. GEMASS, CNRS, Universit\'e de Paris-Sorbonne, France. }\\
\emph{2. CeReFiM and Namur Center for Complex Systems (naXys) Unamur, Belgium} }

\date{\today}

\maketitle
\singlespacing
\begin{abstract}
Understanding the emergence of strong controversial issues in modern societies is a key issue in opinion studies. A commonly diffused idea is the fact 
that the increasing of homophily in social networks, due to the modern ICT, can be a driving force for opinion polariation. In this paper we address the 
problem with a modelling approach following three basic steps. 
We first introduce a network morphogenesis model to reconstruct network structures where 
homophily can be tuned with a parameter. We show that as homophily increases the emergence of marked topological community structures in the networks raises.
Secondly, we perform an opinion dynamics process on homophily dependent networks and we show that, contrary to the common idea, homophily helps 
consensus formation. 
Finally, we introduce a tunable external media pressure and we show that, actually, the combination of homophily and media makes the media effect less effective and leads to 
strongly polarized opinion clusters.  
\end{abstract}

\footnotetext{$^{1}$Correspondence author. E-mail: ygandica@gmail.com, floriana.gargiulo@gmail.com}
\newpage \baselineskip1.0cm
\singlespacing
\section{Introduction}
In modern society we observe the emergence of several controversial issues that can challenge the organization of the society. We have less severe issues, like the diffusion of conspiracy theories and more severe issues, like religious fundamentalisms, that can lead to violent attacks and terrorism. 

In our society, where the communication patterns are so rapidly changing, understanding how these opinion niches are created and reinforced is a key issue: only a full comprehension of these phenomena can suggest the most suitable communication strategies to control or diffuse some ideas. 


Several authors pointed out that a possible responsible of the strong opinion polarization in the society is the particular organization of online social
networks (\cite{polarOSN},\cite{flache3}), enormously enlarging the social pool where the social actors can look for peers (geographical and demographic barriers are broken down), 
allowing people to preferentially enter in contact with people sharing very similar ideas and socio-cultural traits. Moreover, the same filtering algorithms used by the social networks, in order to provide personalized information, amplify this effect favoring the membership in coherent groups and the connection between similar opinions.  
Online social networks amplify the homophily principle, a well known tendency supported by several studies in social psychology (\cite{LAZARSFELD1954}), defined as  the individual tendency to interact preferentially with people perceived as similar. The literature on homophily principle has rapidly  developed in the last years, both with theoretical papers (\cite{birds},\cite{homophily}) and with experimental approaches (\cite{flache},\cite{sociopatterns},\cite{flache2}).

Several authors propose, therefore, that this mechanism of network formation based on homophily directly generates isolated 
echo chambers where the 
information flows remain trapped (\cite{echoChamb1},\cite{echoChamb2}).  
On the other side, all these papers mostly deal with the observation of the opinion flows at a quite initial stage of the 
opinion formation process: using microblogging data we can easily observe how an information spreads among the users, but it
is difficult to track, on long time scales, if and how single users change their opinions.

In this paper we investigate the process of opinion formation in an homophilous environment on the long term, using an ad-hoc simulation framework. Agent-based modeling approach is being broadly used in order to capture the emergent phenomena in opinion dynamics,
when relevant individual mechanisms, postulated by social psychology theories, are applied to several agents 
(\cite{santoRev},\cite{t4}). 

In the context of agent based simulations, several studies on rumor spreading (\cite{abu}) and controversies formation mechanisms (\cite{galam}) have been published in the last decade, but none of these is focused on the social network structure. The basic research question we want to answer here is instead "is homophily in social network a possible origin of the strong opinion polarization observed in our society?". Opinion polarization,  is defined,in our case,  as the emergence of two cohesive opinion groups with radical opinions on a certain subject. Clearly this is not the only possible definition of a so articulated concept. Different definitions of polarization can be found in (\cite{polar}).
 Here we develop our analysis on three levels, gradually extending the complexity of the model. 

First, we focus on the network morphogenesis, where we describe a network generation model, simplifying the implementation proposed in (\cite{Gargiulo2008}), where the local rules for peer selection are based on homophily preferences. \\
We show which are the basic topological properties of these network structures and, in particular, that these local rules, based on opinion, automatically generate the emergence of topological communities in the society. \\
This first result is significant by itself since only few existing models are able to reproduce the emergence of the structural partitions that 
characterize real social networks, using few and essential ingredients. 

Second, we investigate the opinion formation processes on networks displaying homophily. In particular, our interest is oriented to the bounded 
confidence models (\cite{Deffuant2000} and \cite{hk2002}), agent based models, allowing to consider at the same time two central mechanisms of the social 
influence: the tendency toward conformity - explained by social comparison theory (\cite{Festinger1957}) and social balance theory (\cite{socbal1},\cite{socbal2})-  and confirmation bias - the
tendency to filter out informations that are too far from our points of view. In particular, we will consider the Deffuant BC model (DW), where the 
opinion evolution is based on peer interactions. 

It has been shown, in (\cite{fortunato2004}), that the outcomes of the bounded confidence models are topology independent on static networks. On the contrary, in (\cite{groupBased},\cite{groupBased2}, \cite{abu}, \cite{kozma}) the authors showed that the results can strongly change on (co)-evolving networks. 
Here we address the key question about the connection between homophily and opinion polarization (\cite{NEUMANN2013}): is homophily promoting opinion controversies in the society as evoked, for example, in (\cite{echoChamb1},\cite{echoChamb2})? \\
We show that, on the stable final configurations, the contrary is true: networking based on homophily promotes consensus, due to the interplay of the dynamics inside and between the community structures.

Third, we add in the simulation framework a further ingredient represented by the traditional media, spreading with a more or less marked pressure, the opinion of the society's empowerment. 
To model the media we extend the bounded confidence framework to an asymmetrical interaction between human agents and media, considering  that the confirmation bias in the media exposure is a well documented fact, usually called selective exposure (\cite{selectiveExp}).
In every society, although the dominant tendencies usually follow the message promoted by dominant institutions (i.e. corporations for mass media, religion and educational institutions) some other less dominant tendencies/opinions always appear. A lot of effort has been devoted in trying to understand the mechanism leading to this evident opinion diversification. 
A counterintuitive effect, regarding the effect of mass media, has been observed in BC models.  It has been shown in several papers  (\cite{Carletti2006},\cite{mediaGLM},  and \cite{Pineda2015}) that, if the media pressure is low, media are able to attract all the opinions but, on the contrary, oppressive propaganda gives rise to the emergence of opposite extreme opinions both in the Deffuant model (DW) and  in the Hegselmann and Krause (HK) model. 
A similar effect has been also observed in the Axelrod's model for the dissemination of culture, in (\cite{Avella2005}, \cite{Gandica2011}) and in particular in this recent study (\cite{mediaAxel}) where, coherently to our case, the interplay between word-of-mouth and media is considered. 
This over-exposure phenomenon is  well known in marketing studies (\cite{Groucutt2004}).

Although all these studies address the research question of the formation of counter-message competing with the dominant mass media, in this work we explicitly focus on the structure of the non-aligned states. In (\cite{Carletti2006},\cite{mediaGLM},  and \cite{Pineda2015}) it has been shown that, when the interaction is constrained by non-homophilous connectivity, the final outcome of  the model, for high media pressure, is a strong cluster aligned with the media and a large number of unclustered opponent opinions. A second cohesive counter opinion cluster cannot be formed in BC models over regular complex networks. Here we show that, on the contrary, the presence of homophily in social networks over a media dominated system, plays a central role in the recomposition of a strong opponent cluster, leading to the final polarization of the opinions. At the same time we show that homophilous system are much more robust to media propaganda, allowing the formation of counter-clusters also for lower values of the media pressure. 

Differently by the co-evolution frameworks (\cite{groupBased},\cite{groupBased2}, \cite{abu}, \cite{kozma}) where opinions are
updated together withe the network structure, in this paper we consider that network morphogenesis and opinion dynamics take 
place at different time scales. The formation of social network is a slow process and, in this case, the \emph{opinion} on
which we base the homophily choices, is an abstract representation of a global vision of the world. Opinion dynamics processes
represent the formation of a global opinion on a concrete subject (a new law, a referendum, a piece of news, etc). This
processes have a fast dynamics that do not allow to the networks to coherently reshape. In this case, what we define opinion 
is the particular judgement that an individual, with a certain vision of the world, has on this subject. In this sense, the representation of the opinion in these processes is a sort of local characterization of the opinion on which network morphogenesis is based. 

The paper is organized as follows: In section 2 we present the network morphogenesis model and the topological properties of the obtained networks. In 
section 3 we present the opinion evolution on homophily-based networks. In section 4 we show the combined effect of media propaganda and homophily. Conclusions
are presented in section 5.


\section{The network morphogenesis}
\subsection{The Model}
We first propose a growing network model allowing to fine tune the  homophily level. The growing network approach for network morphogenesis is a dynamical process where at each time step new nodes and new links enter the network and/or old links are canceled or rewired. We will consider the simplest case where at each time step a single node is added to the network with a set of associated links to the pre-existing nodes.

The probability that the new node $N$ gets connected to the pre-existing node $i$, $\Pi_{N\rightarrow i}\sim\varphi_N(i)$ contains the selected mechanism for network growth. The fitness function $\varphi_N(i)$, associated to each pre-existing node,  represents how attractive is a pre-existing node $i$, for the new node $N$, to establish a link. \\

It is well known that a fitness function based on the degree ($k_i$) of the pre-existing nodes $\varphi_N(i)= k_i$, namely a situation where a node with a large connectivity (measured as its degree) has a larger chance to attract new links, gives the preferential attachment mechanism generating scale free networks with a power law degree distribution, hereby named BA-networks, (\cite{Barabasi2007}).

Implementing a fitness function based on the degree is motivated by the larger \emph{visibility} that highly connected nodes have: more friends I have, more probable is that I am present in different social circles and more probable is to meet new friends. Moreover this is one of the mechanism on which friendship recommendations in online social networks are based. 

To include homopily in the morphogenesis without forgetting the connectivity issue, in our setup,  we construct the fitness function $\varphi_N(i)$ so that the new
node has a preference to get connected both to high degree nodes and with nodes with similar opinion:
\begin{equation}
\varphi_N(i)= k_i \exp (- \beta | \theta_N - \theta_i| ),
\label{fitness}
\end{equation}
where $k_i$ is the degree of the ancient node $i$, $\theta_i$ its opinion, $\theta_N$ the opinion of the new node and $\beta$ a coefficient tuning the homophily effect.

Let us note that when $\beta=0$ the growing mechanism follows the classical preferential attachment, leading to the
usual Barabasi-Albert network, with  power-law degree distribution $P(k)\sim k^{-3}$. When $\beta \neq 0$ the homophily 
comes into play and a competition between the preferential degree attachment and the opinion similarity takes place. 
In the limit $\beta \rightarrow \infty$ only homoplily matters.

The model follows the following steps:
\begin{itemize}
\item Each node is initialized with an opinion randomly selected in a continuos interval between  $\theta_i\in [-1,1]$. 
\item The network generation process starts from an initial fully connected structure with $N_{ini}=5$ nodes. 
\item At each step a new agent, $N$, enters in the network
\item The new agent $N$ gets connected to $m$ 
pre-existing agents ($m$ new links) using a roulette-wheel selection process (or fitness proportional selection), based on the probabilities:
\begin{equation}
\Pi_{N\rightarrow i}=\frac{\varphi_N(i)}{\sum_{i=0}^N\varphi_N(i)}
\end{equation}
Notice that the parameter $m$, namely the number of new links added to each new node has no influence on the global properties of the network (like degree distribution, clustering, mixing, etc.). This parameter defines the minimum degree of the network and the total number of edges.\\
\end{itemize}

\begin{figure}[!h]
\centering
\includegraphics[width = 16cm]{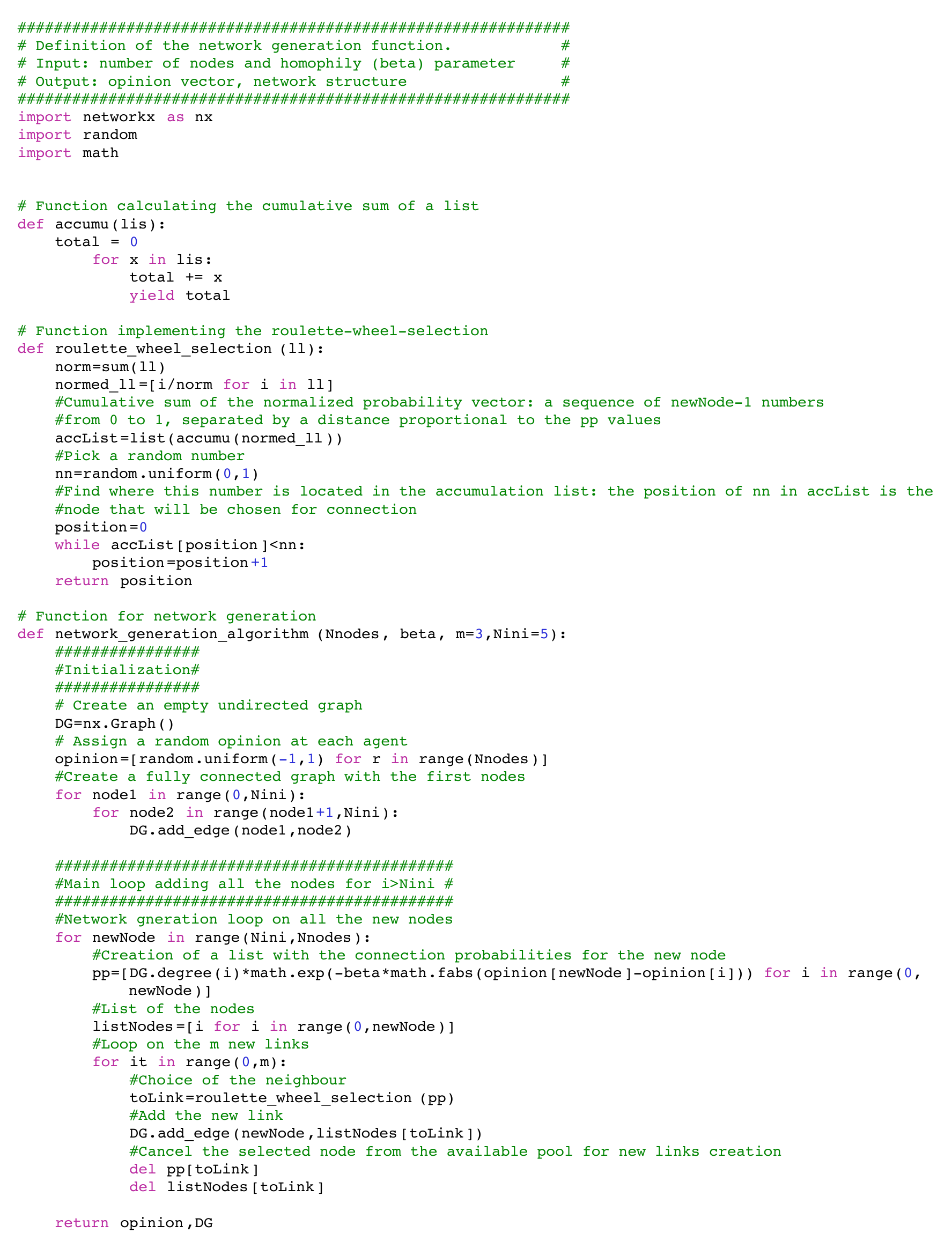}
\caption{Python function to generate networks.}
\label{alg:1}
\end{figure}

In Fig.\ref{alg:1} we report the the Python implementation of the algorithm for the network morphogenesis.

\subsubsection{Results}
\begin{figure}[!h]
\centering
\includegraphics[width = 16cm]{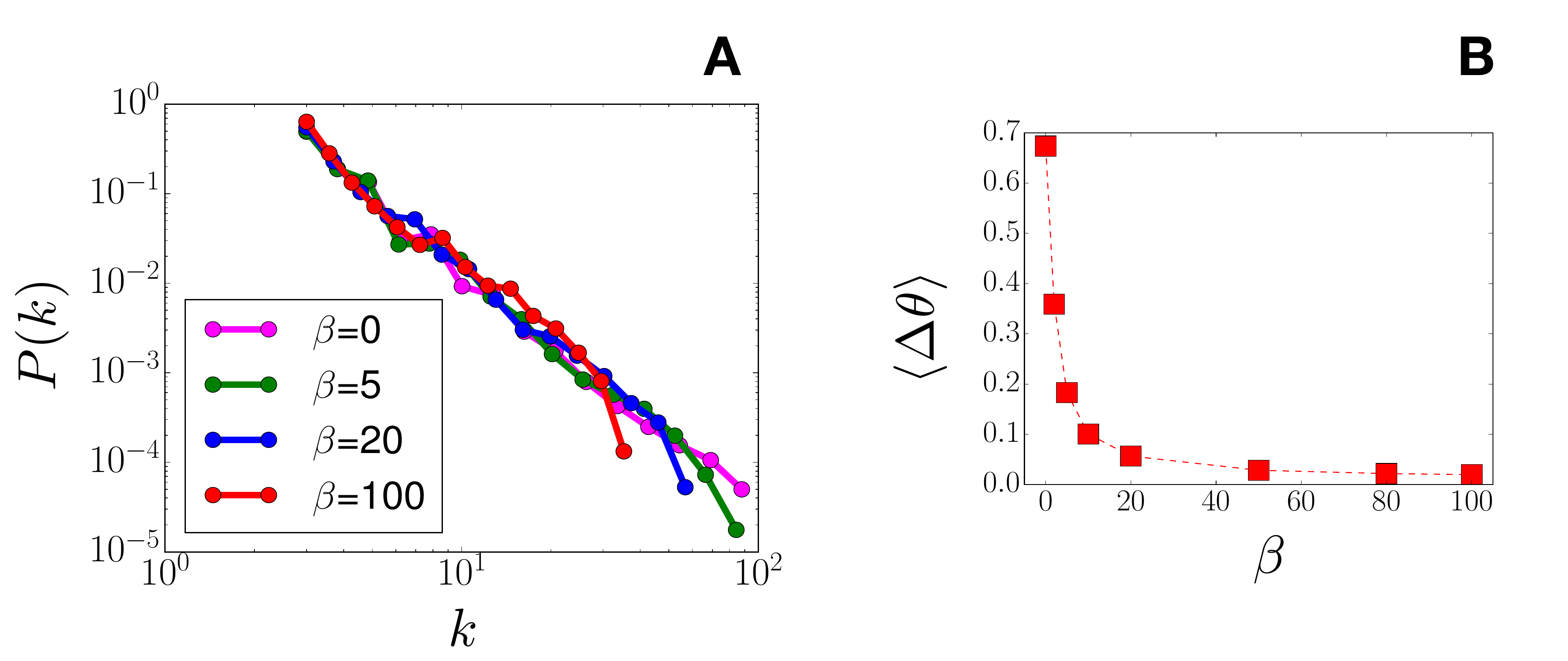}
\caption{A) Power-law degree distribution for several values of $\beta$ over Scale Free (SF) networks with $N=1000$,
$m=3$ and $N_{ini}$ (initial core) $=5$. The homophily mechanism ($\beta \neq 0$) leads to structures following the same power-law 
degree distribution but with a cutoff, imposing a maximum value of degree by increasing $\beta$. B) Average opinion distance between 
connected nodes. Connection between more similar nodes is clearly increasing with the homophily parameter}
\label{fig:1}
\end{figure}

Due to the presence of the connectivity in the fitness function $\varphi_N(i)$ (Eq. \ref{fitness}), the final degrees of the network are distributed for a large range, as in the original BA model ($\beta=0$), on a power law distribution. 
Increasing $\beta$, the homophily mechanism leads to a
cutoff in the distribution, decreasing the maximum value of degree  (figure \ref{fig:1}-A). 

In figure \ref{fig:1}-B we show the average opinion distance between connected nodes. The opinion similarity between connected
nodes increases extremely fast as the homopily parameter $\beta$  is switched on.


\begin{figure}[!h]
\centering
\includegraphics[width = 16cm]{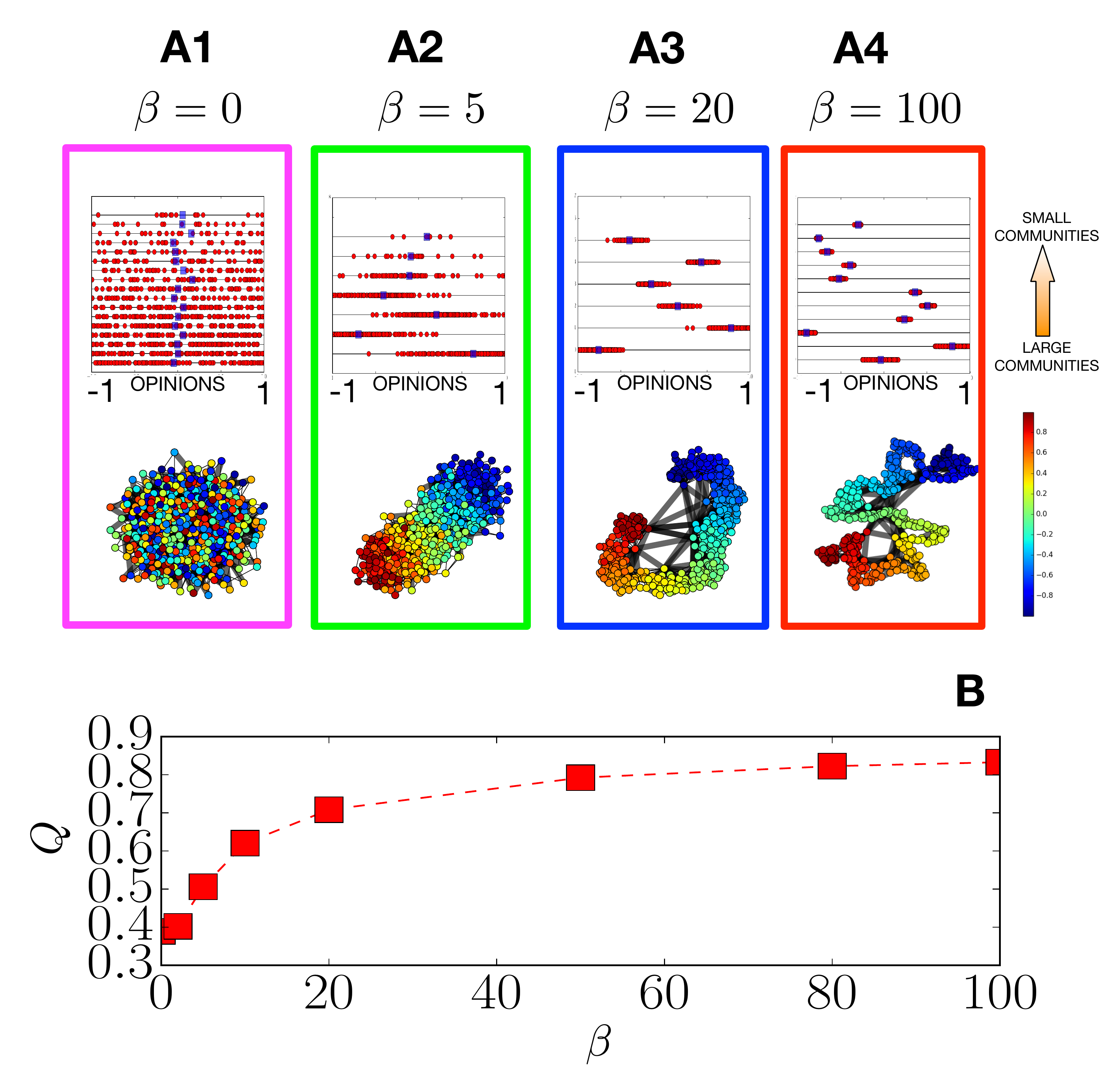}
\caption{A1-4) Upper plot: each line represent a community (ordered from the larger on the bottom to the smaller on the top); the red points represent the agents opinions inside each community, the blue square is the average opinion of the community. Increasing $\beta$ the opinion range inside the communities is smaller. Lower plot: network visualization using a force layout (allowing to visualize the partitions). The color of the nodes depends on their opinion.
B) Community modularity as a function of $\beta$. The results are the average values of 100 replicas of the morphogenesis process, for a network 
with $N=1000$ and $m=3$. Modulatity increases with $\beta$, meaning that more significant community structures are formed.}
\label{fig:2}
\end{figure}

The results obtained in Fig.~\ref{fig:1} are the direct consequence of the preference function structure. A more relevant 
emergent property can be observed in figure \ref{fig:2}: networks with high homophily exhibit meaningful community structure. 
Topological communities are groups of nodes that are more strongly connected among them than with the rest of the network. 
Notice, however, that there is a strong difference between community structures and network disconnected components: communities
are largely connected among them but links exist also between the communities, connected components are totally disconnected 
among them. Several real network  structure present signature of this properties - citation networks, mobility networks, social networks, 
semantic networks.. (\cite{newman}).  At the same time few network morphogenesis models are able to reproduce these patterns (\cite{bianconi}).

Several algorithm exist to identify community structures For a review see for example (\cite{santoComm}). In the following we 
used the Louvain algorithm (\cite{louvain}, http://perso.crans.org/aynaud/communities/. 

The goodness of a partition is measured by the modularity ($Q$), a function comparing the concentration of edges within communities, in the network, with a rewired network with a random distribution of links (obtained with the configuration model). 
Large values of the modularity ($Q\rightarrow 1$) signify that the community structure is highly significant, namely the fraction of links within the community largely exceed the fraction of links between the same nodes in a random configuration.
For a mathematical definition of the modularity see (\cite{modularity}) 

The best partition is the configuration that maximize modularity. At the same time, if the community structure is not a significant marker of a network, modularity remains low also for the best partition. 
In figure \ref{fig:2} we display the modularity measure for the best partition of the network, as a function of the homophily parameter $\beta$. Modularity monotonically increases with $\beta$, implying that the presence of communities is  a natural emergent effect of the homophily preference in network morphogenesis.

In the lower plots A1-4 of figure \ref{fig:2} we present a network visualization for different values of the homophily parameter. The used visualization layout (based on a force algorithm) has a repulsive force to push away disconnected nodes and a spring-like force attracting connected node. The result of this visualization algorithm is the spatial separation of the network communities: when the community structure is significant few connections (represented by the large black lines in the plot) exist between the communities that will be therefore pushed away among them, while the large number of links inside the communities will spatially group the nodes of the same community. We can observe that for the BA network ($\beta=0$) communities are not visible, while a more and more structured shape appear as homophily is switched on. Notice that the communities are largely uniform in term of opinion. At the same time links exist between all the communities. This factor is central for understanding the opinion propagation dynamics that will be described in the next section. 

In the upper plots A1-4 of figure \ref{fig:2} we show the agents opinions inside each community. Each line in the plots represents a community, ranked from the largest (on the bottom) to the smaller (on the top). We observe that for the BA network the opinions are randomly distributed in all the communities and the average opinions of the communities coincide with the center of the opinion interval. When the homophily is present the communities \emph{specialize}: their average opinion moves from the center. Largest it the value of $\beta$ smaller is the dispersion of the opinions around the average opinion of the community. 

%
%

In all our experiments the initial opinion has been initialized according to a uniform distribution. We tested that the aggregated results we presented in Fig.~\ref{fig:1} and Fig.~\ref{fig:2}B are robust to a changing in the simulation paradigm: before fixing the same opinion vector and after building different networks on this opinions. 
Notice that using different distributions for the initialization (like i.e. a Gaussian) could change the final outcome of the process. Exploring this issue is out from the scopes of this paper (addressed mostly to understand the relationship between homophily and opinion propagation) and we leave this direction open for subsequent studies.

\subsubsection*{ Central result:} Homoplily as ingredient for network morphogenesis leads to networks where the community 
structure is a fundamental marker.

\section{Opinion dynamics}
\subsection{The model}
In the previous section we described a morphogenesis algorithm to build network structures based on homophily. In this section we will show how this network structure influences opinion dynamics processes. 

Several different mechanisms can drive opinion formation. In peer interactions two main factors have been observed as fundamental forces for mutual influence. The first on is \emph{conformity}, a mechanism due to the psychological need to reduce conflict among peers that consists in the reduction of opinion distances after an opinion exchange. The second one is the \emph{confirmation bias} that is the selective filtering of opinions too far from ours. 
A well known model taking into account both these mechanisms is the bounded confidence (BC) model (\cite{Deffuant2000}). 
This model depends on a tolerance parameter $\varepsilon$ tuning the importance of confirmation bias. According to this model, once two agents $(i,j)$ are selected as peer for an interactions, they will update their opinion using the following threshold rule:
\begin{equation}
if |\theta_i-\theta_j|<\varepsilon\Rightarrow\begin{cases}
               \theta_i=\theta_i+\mu (\theta_j-\theta_i)\\
               \theta_j=\theta_j+\mu (\theta_i-\theta_j)
            \end{cases}
\label{interact}
\end{equation}

If the velocity parameter is $\mu=1/2$, as it is usually fixed, the two opinions will converge to their average.  

\begin{figure}[!h]
\centering
\includegraphics[width = 14cm]{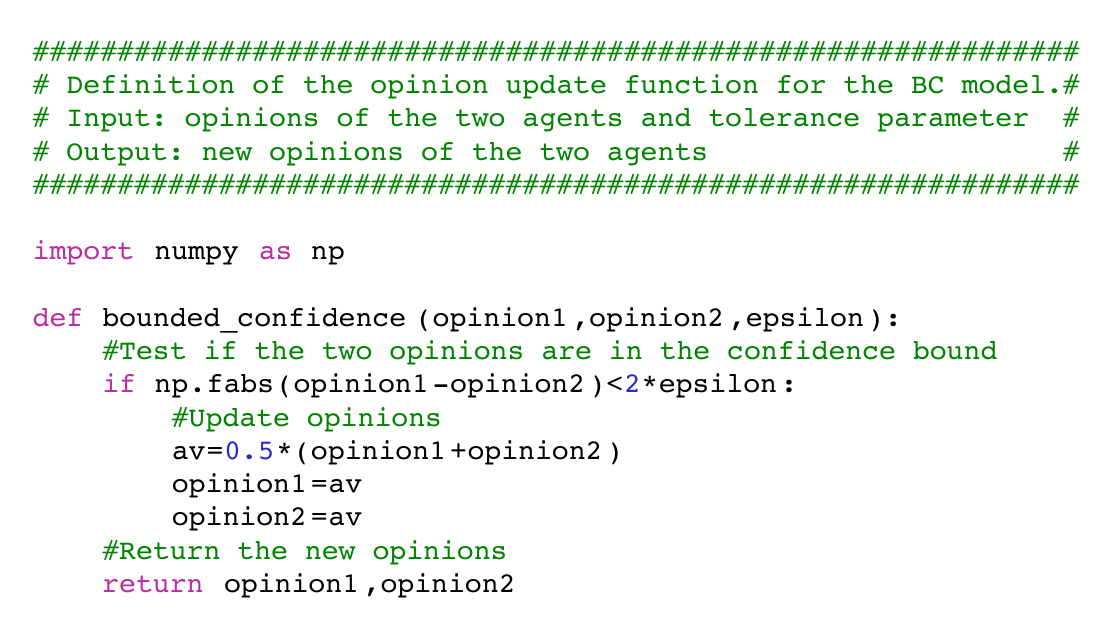}
\caption{Python function for a single update of the BC model.}
\label{alg:2}
\end{figure}

The Python implementation of the function for a single interaction with the BC model is presented in Fig.~\ref{alg:2}.

It is well known that, independently from the choice of the parameter $\mu$ the repeated application of this rule, on random 
pairs of peers in a fully connected population, drives to two possible scenarios: \emph{total consensus}, where all the 
opinions converge to the initial opinion average, for $\varepsilon\geq 0.5$ and \emph{opinion clustering}, where two or more
opinions coexist, for $\varepsilon< 0.5$ (\cite{ian})

In the original description of the BC model (\cite{Deffuant2000}), all the agents can interact with all the others. This complete mixing assumption can be quite unrealistic when we consider a large number of agents that cannot be in contact with anyone else in the society. People can interact and exchange opinions only with the peers that they meet in their everyday life (online or offline), namely with the neighbors in their social network. A second important step, after (\cite{Deffuant2000}), is therefore to constrain the peer selection only to couples of nodes connected by a link in a network structure.

It has been shown in (\cite{fortunato2004}) that the same scenarios and the same transition threshold to consensus $\varepsilon_c=0.5$ is observed if the peers are selected only between the edges of a more static complex network structure (random graphs, small world networks, scale free networks). 
On the other side, it has been shown in (\cite{kozma}, \cite{groupBased},\cite{groupBased2})
that the consensus threshold can change on dynamical networks, once some rewiring based on nodes' opinion can take place. 

In the following we will study how the parameter $\beta$, tuning the homophily level in networks, can influence the outcome of the BC model. Namely we will address the question: Does a larger homophily in the network structure implies the formation of opinion bubbles?

\begin{figure}[!h]
\centering
\includegraphics[width = 16cm]{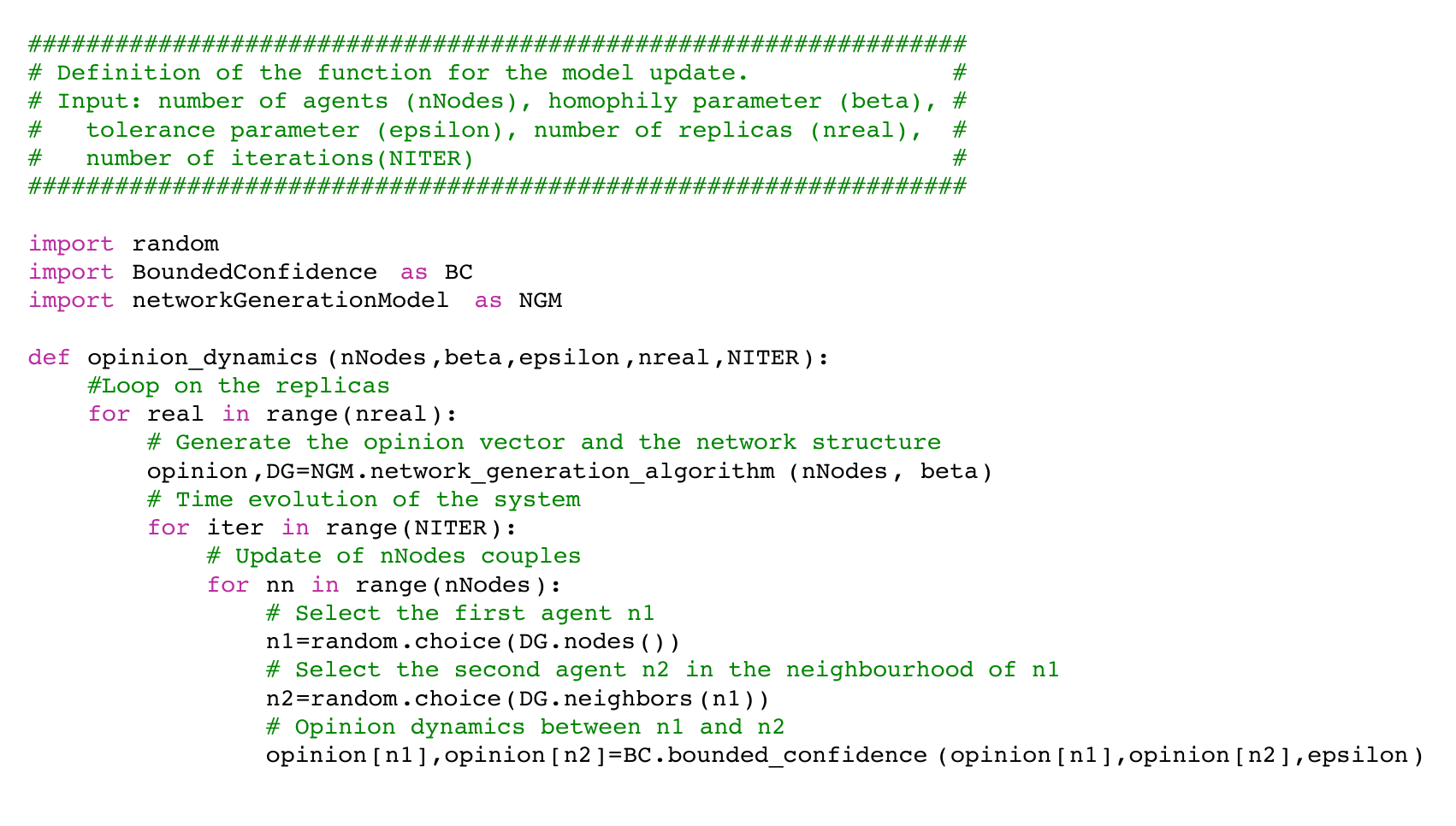}
\caption{Python implementation of the opinion dynamics model.}
\label{alg:3}
\end{figure}

The model evolves according to the following steps (see Fig.~\ref{alg:3} for a Python implementation):
\begin{itemize}
\item A random uniform opinion distribution and a network structure (with $N$ nodes and with a $\beta$ parameter) are created.
\item At each time step $N$ couples of neighboring nodes (connected by a link of the network) are selected and a pairwise interaction with BC model is performed (asynchronous update).
\item The loop is halted when, on all the edges no more successful interactions are possible ($\|\theta_i-\theta_j|>\varepsilon$ or $\|\theta_i-\theta_j|=0$)
\end{itemize}

\subsection{Results}
\begin{figure}[!h]
\centering
\includegraphics[width = 15cm]{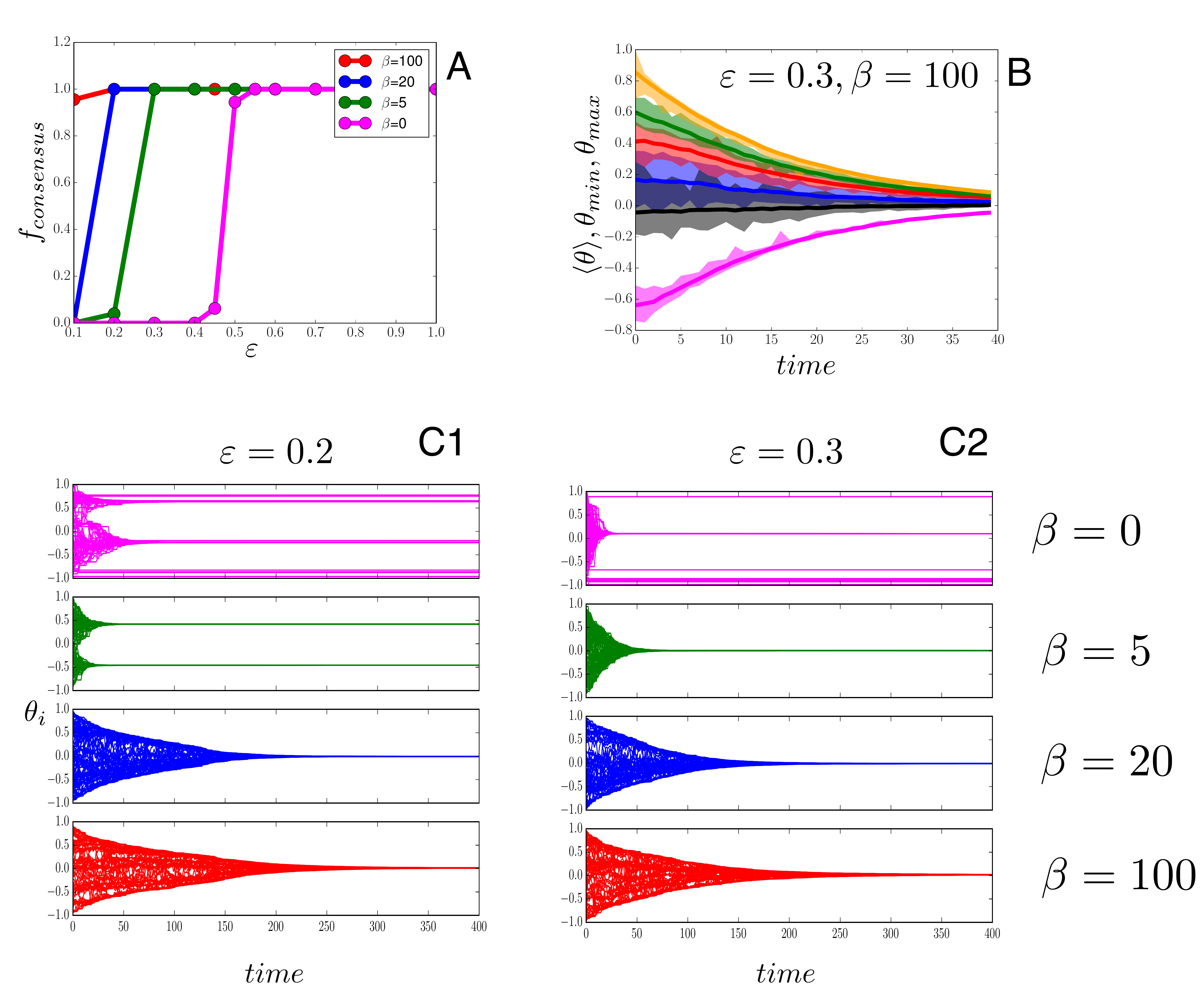}
\caption{A) Fraction of replicas leading the system to consensus as a function of $\varepsilon$. For each value of $\beta=0,5,20,100$, the simulation has been performed $N_{replicas}=100$ times. Consensus is reached for lower values of $\beta$ for homophilous networks. B) Each color represent the opinion span and the average opinion of a community for a single replica of a system with $\varepsilon=0.3$ and $\beta=100$. There is an interplay between the fast dynamics inside each community and the slower dynamics between the communities. C1,2) Single replica representation of the evolution of the individual opinions (each line represents the opinion of an agent), for $C1\rightarrow\varepsilon=0.2$ and $C2\rightarrow\varepsilon=0.3$, and for different values of $\beta=0,5,20,100$}
\label{fig:4}
\end{figure}

In figure \ref{fig:4} we show that the answer to the question  \emph{"Does a larger homophily in the network structure implies the formation of opinion bubbles?"}, is clearly negative and that, on the opposite, a large homophily reduce the existence of radical issues. 

In figure  \ref{fig:4}A we plot the fraction of replicas ending up to consensus for different values of the homophily parameter. For $\beta=0$ (Barabasi-Albert network), we observe the transition at $\varepsilon_c=0.5$ predicted in \cite{fortunato2004}. 
But for higher values of $\beta$ we observe that the transition threshold becomes smaller and smaller ($\lim_{\beta\rightarrow\infty}\varepsilon_c(\beta)=0$), meaning that the system will always end up to consensus. 

Figure \ref{fig:4}C shows how it happens. Remember that the system can evolve until on some links of the social network the agents connected by the link have two different opinions at a distance smaller than the tolerance. At the moment when, on all the links the agents have the same opinion or their opinion difference is larger than $2\varepsilon$, whatever pair is selected for the opinion dynamics, opinions will not change anymore. 
 
For low values of $\beta$ the system get frozen at a very initial phase. After few iterations the agents cannot find in their neighborhood any peers with whom having "positive exchanges" (for all the couples $|\theta_i-\theta_j|\geq 2\varepsilon$). The agents with a moderate opinion rapidly converge forming a major cluster (located around the average opinion of the system $\langle\theta\rangle=0$). Larger is $\varepsilon$, larger is the central cluster size. Since the link construction is independent by the opinion, there is therefore a large probability that the radical agents are connected with agents in the majoritarian cluster (positioned at an opinion distance larger than $2\varepsilon$ from their actual opinion) and not among them. The radical agents, remain therefore isolated, keeping their initial opinion. 

For larger values of the homophily $\beta$ the dynamics is slower, but much more "inclusive". Since radicals agents are now mostly connected with similar, they do not remain isolated. Agents always find peers with an opinion sufficiently near to interact, and therefore their opinions change gradually at each interaction. 

As we can observe in Figure \ref{fig:4}B (where each color represents the opinion span in each community) the dynamics happens at two levels: a rapid convergence inside the communities and a slower one between the communities. The large number of links inside the communities allow the fast convergence to the average opinion of the communities, at the same time the links between the communities (in their turn connecting communities with similar average opinions) allow a slow dynamics of the average opinions of the communities, toward consensus. \\
To use a visual conceptualization, the homophily structure, provides a sort of continuous path allowing the radical opinions to join the central ones. 

\subsubsection*{ Central result:}
In a situation where opinion evolve in time, homophily in social networks favors consensus formation. Therefore, contrarily to the common idea that online social networks are directly responsible for the installation of sever opinion wars in the society, we can argue that, on larger time scales, homophily helps the resilience of the society to the presence of radical issues. 

\section{Opinion dynamics in the presence of dominant media}
In this section we add a further ingredient in the system: the presence of an external media, diffusing with a certain pressure $p_m$ a constant opinion $\theta_M$. 

In several previous papers it has been observed, in the context of BC models, that the dominant media lose their capacity to attract people opinion, (i.e., the effectiveness of propaganda) after a certain pressure threshold.
In \cite{Pineda2015} the external media pressure has been modeled as a heterogeneous open mindedness distribution and some
interesting particularities are reported, due to the specific conditions considered. 
In Carletti et al. (\cite{Carletti2006}) the mass media has been modeled as a periodic perturbation. In this paper the authors divided the systems response into four regimes, where the efficiency of the message is explained in terms of the  people open-mindedness threshold, ($\varepsilon$), and the period of the message. In 
this work the authors stress the importance of the collapse into clusters before the exposure to propaganda, given the influential role that community 
structures can develops to profile the opinions around a message. This phenomenon is a natural connection with our work, where the effect of the dominant message faces a strong community interaction.

Opinion dynamics with media is an asymmetrical opinion update, meaning that, after an agent interacts with media, she can change her opinion, while the opinion of the media ($\theta_M$) will remain identical. 
We extend to this asymmetrical framework the structure of the BC model:
\begin{equation}
if |\theta_i-\theta_M|<2\varepsilon\Rightarrow\begin{cases}
               \theta_i=\theta_i+\mu (\theta_M-\theta_i)\\
               \theta_M=\theta_M
            \end{cases}
\label{interactMedia}
\end{equation}

\begin{figure}[!h]
\centering
\includegraphics[width = 14cm]{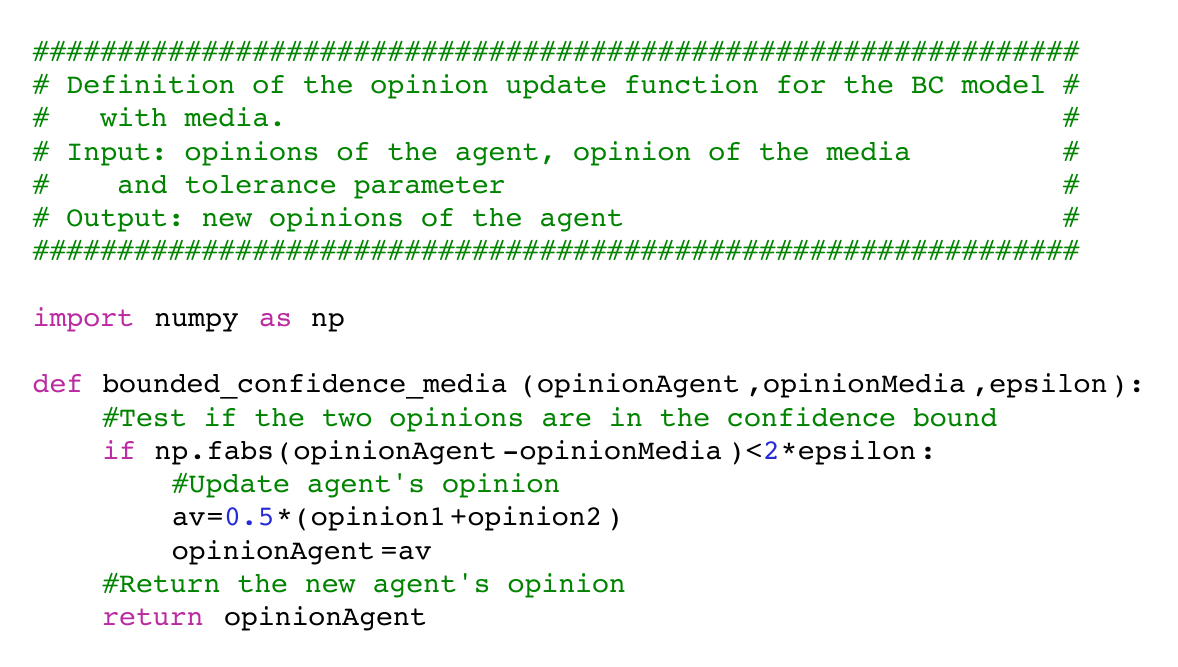}
\caption{Python implementation of the BC interaction between an agent and media}
\label{alg:4}
\end{figure}

In the following we will fix $\mu=0.5$, as in the previous case, and the opinion of the media, $\theta_M=1$. The Python function defining this
asymmetrical opinion update is described in Fig.~\ref{alg:4}

To analyze the effect we introduce a parameter $p_m$ representing the exposure to dominant media messages. If $p_m=0$ the agents have no probability to 
interact with the media, while if $p_m=1$ the agents will interact only with the media. 

\begin{figure}[!h]
\centering
\includegraphics[width = 18cm]{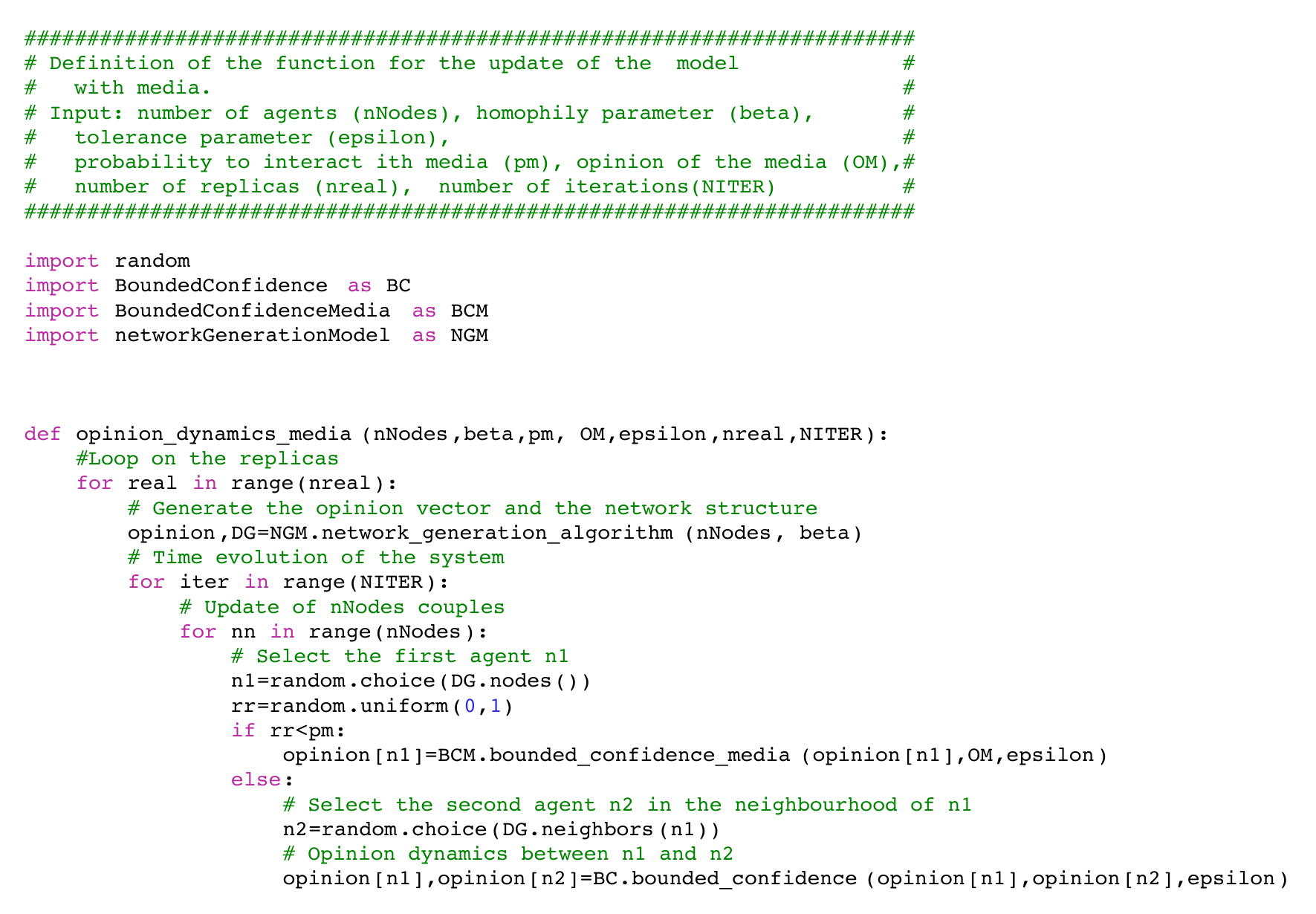}
\caption{Python implementation of the opinion dynamics model with media.}
\label{alg:5}
\end{figure}

The model evolves according to the following steps (for a Python implementation see Fig.~\ref{alg:5}):
\begin{itemize}
\item A random uniform opinion distribution and a network structure (with $N$ nodes and with a $\beta$ parameter) are created.
\item At each time step, for $N$ times, an agent $i$ and a real number in the interval $r\in [0,1]$ are randomly extracted.
\item If $r<p_m$: agent $i$ makes opinion dynamics with the media according to Eq.\ref{interactMedia}
\item If $r\geq p_m$: a second agent $j$ is selected and an opinion dynamics update according to Eq.~\ref{interact} is performd on the pair ($i,j$).
\item The loop is halted when, on all the edges no more successful interactions are possible ($|\theta_i-\theta_j|\geq2\varepsilon$ or $|\theta_i-\theta_j|=0$)
\end{itemize}

In order to compare the same scenarios according to the unforced opinion dynamics, first we will consider the case $\varepsilon=0.5$ where, without forcing, consensus is obtained both for homophilous ($\beta=30$) and non-homophilous ($\beta=0$) situations. In a second step we will extend these results to a larger spectrum of values for the tolerance threshold.

\subsubsection{Results}
\begin{figure}[!h]
\centering
\includegraphics[width = 16cm]{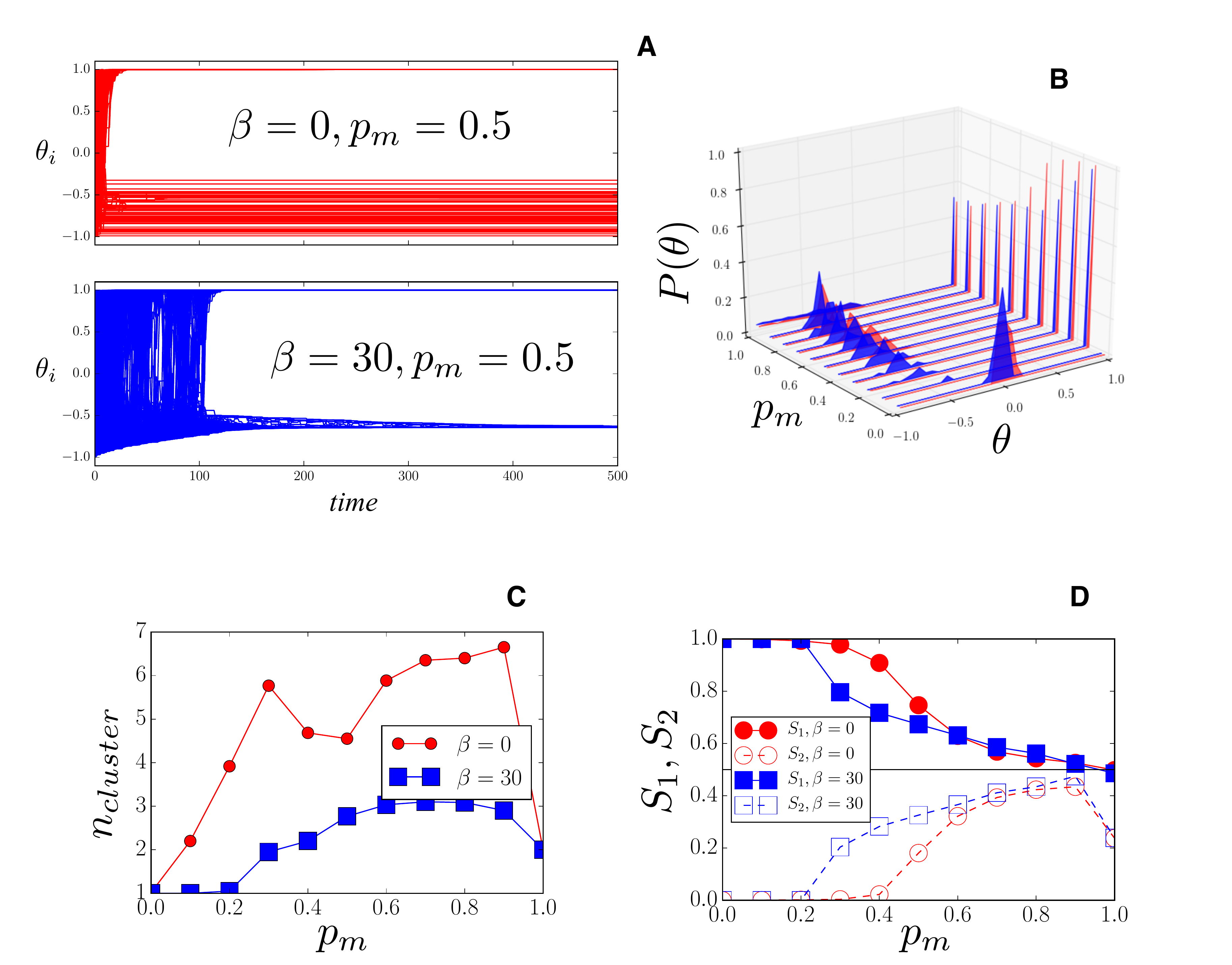}
\caption{A) Single replica evolution of the system for $\varepsilon=0.5, p_m=0.5, N=1000$, for two values of the homophily parameter: $\beta=(0,30)$. B) Final state representation for 20 replicas of the system and for different values of the media pressure. As $p_m$ increases the cluster aligned with media loses mass. C) Average number of clusters as a function of the media pressure $p_m$ for two values of the homophily parameter: $\beta=0,30$ and for 20 replicas of the system. When homophily is present less clusters are formed. D) Average size of the first (filled markers) and second (empty markers) clusters as a function of the media pressure $p_m$ for two values of the homophily parameter: $\beta=(0,30)$ and for 20 replicas of the system. When homophily is present the secondary cluster becomes larger.}
\label{fig:5}
\end{figure}

Our results are reported in figure ~\ref{fig:5}. In figure ~\ref{fig:5}A we show the result of the evolution of a single replica in the case where the media pressure is fixed to $p_m=0.5$. 
When agents are highly affected by mass-media, like in this case ($p_m=0.5$), several opinions remains in opposition to the dominant message. This effect
has been observed over BC models in all the previously cited papers (\cite{Pineda2015},\cite{Carletti2006},\cite{mediaGLM}) as well as in the
Axelrod model (\cite{Gandica2011}).

Notice, however, (Fig.~\ref{fig:5}A) that when homophily is not present in the social network, these counter-messages remain a non-homogeneous set of separate opinions (the parallel red lines in the figure) because the agents carrying these opinions are not connected by the social network and therefore cannot interact among them. The presence of homophily in the social network, makes more probable that the individuals with radical opinions opposing to the media cluster, are connected among them by the social network. The interactions between these radical agents are therefore possible allowing the  recomposition of a secondary composite opinion counter-cluster (the final blue state in the figure), describing a real situation of opinion polarization. \\

In the following we analyze this effect more deeply as the result of aggregate replicas of the model and for various values of the media pressure parameter $p_m$.
In figure ~\ref{fig:5}B is shown the probability density function for the final states in terms of 
mass-media intensity. The red shapes represent the case for $\beta = 0$, the blue shapes represent the case for $\beta = 30$, where homophily strongly influence the network structure. \\
In figure ~\ref{fig:5}C and D we respectively display the number of opinion clusters and the size of the two largest clusters as a function of the media pressure, $p_m$. 

In general we observe that the alignment of all the agents to the mass media state (an unique opinion cluster at $\theta=\theta_M$) occurs, counterintuitively, only for low values of the media intensity, as previously found in several works, and explained in the introduction. This happens because the fast drift toward media opinion leaves several isolated agents that cannot find a peer to interact with. For networks without homophily, the threshold value for the mass-media to have this 
self-defeating effect is around $p_m = 0.4$, when some very small non-aligned states start to appear (Fig. ~\ref{fig:5}C), and become macroscopic (Fig. ~\ref{fig:5}D). \\
A first effect of the presence of homophily in network morphogenesis is the lowering of the threshold for the media to be 
effective. In the case where homophily is present we observe the formation of counter messages (and the decrease of the size 
of the cluster aligned with the media) already for $p_m=0.2$ (Fig.~\ref{fig:5}B). 


At the same time, if we look at the blue plots in figure ~\ref{fig:5}B, for high values of $p_m$, it is clear how the homophily-based networking structures the non-aligned states around one powerful cluster (as we observed for the single replica plot in Fig~\ref{fig:5}A) . The fact that the cluster has a larger amplitude than the single peak observed in (Fig.~\ref{fig:5}A) is due to the statistical fluctuations of the position of the second cluster among the different replicas of the system (for each replica the single cluster observed in Fig.~\ref{fig:5}A  has a different position), but we can observe,  as well, that the final opinions are much less sparse than in the case without homophily. \\
This effect can be better observed in Fig. \ref{fig:5}C and D. The community structure resulting from the homophily effect during the network morphogenesis, re-organizes all the small non-aligned states into a macroscopic one, competing with the dominant message: for the homophilous case, for  $p_m \geq 0.4$ the size of the second cluster is much larger  while the number of cluster is much smaller.
At the same time we can observe a relevant reduction of the size of the dominant cluster (aligned to the media).

\begin{figure}[!h]
\centering
\includegraphics[width = 16cm]{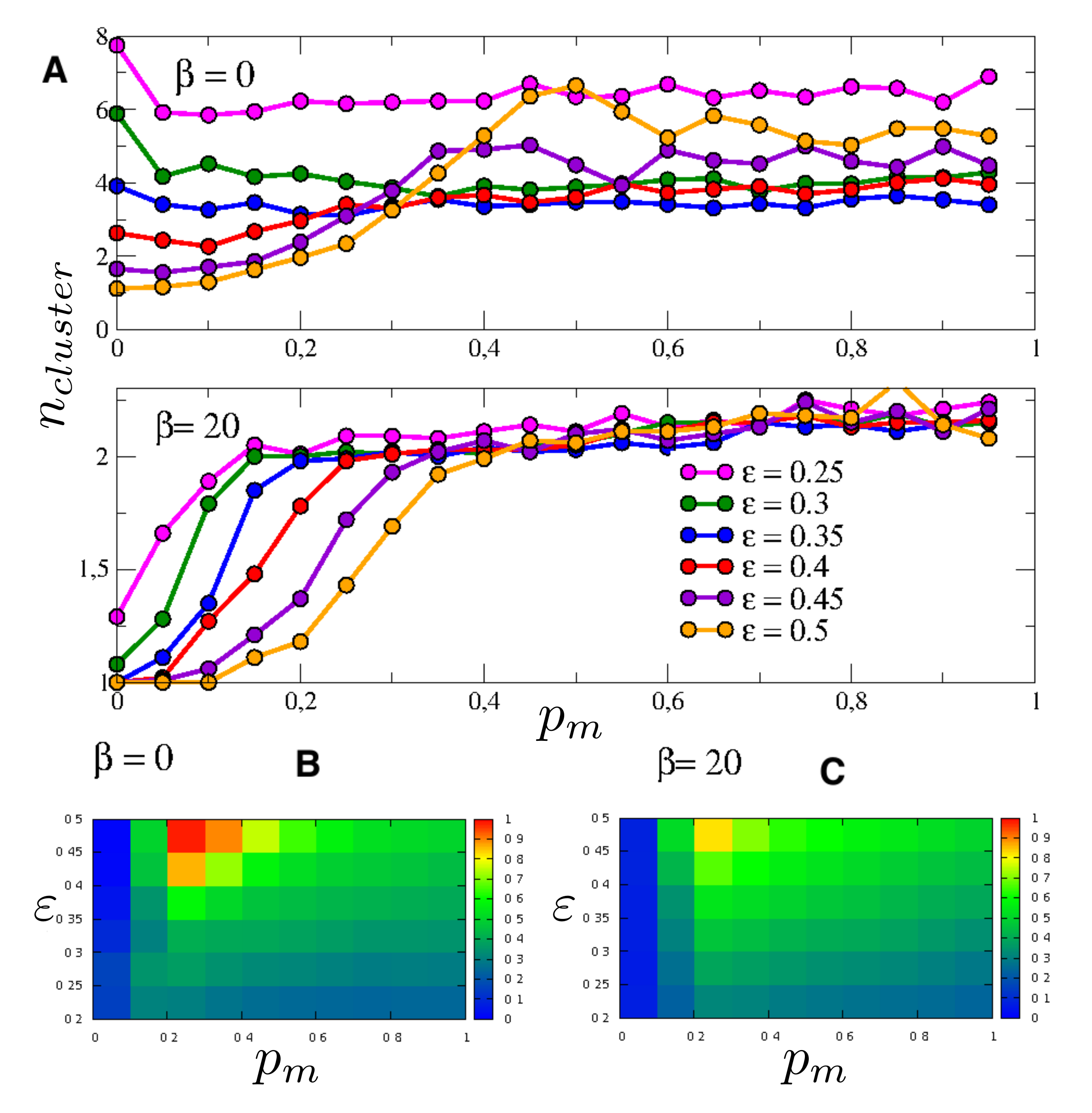}
\caption{Upper Panel (A) : Number of clusters as a function of the media pressure, $pm$, for two values of the homophily parameter: $\beta = 0, 20$ and 
six values of $\varepsilon$. Each point is the average over $100$ replicas. Bottom panels: Averaged proportion of agents aligned with the imposed mass \
media, for  $\beta = 0$ (in B) and $20$ (in C). Results in B and C show how the strength of community structures undermines the dominant 
role of mass media reported in several ABM's. On the other hand, results in A (upper panel) show how the community structure
re-organizes all the possible different opinions in one counter-message position, independent of the general tolerance of the 
society. However, the lower the tolerance, the faster the counter message appears.}
\label{fig:6}
\end{figure}

In Fig. \ref{fig:6} we extend the analysis to other values of the tolerance threshold $\varepsilon$. Notice however that in these cases the output of the model is a priori different already at the level of the opinion dynamics without media. The number of clusters (in upper panel) and the fraction of agents aligned with the external
mass-media (in the bottom panel) are shown, varying the media pressure $p_m$ but also for several values of tolerance. 

This global picture first shows that, as in the case where media are not present, when the tolerance is low more clusters are formed (\ref{fig:6}A). This effect  is more pronounced for the case where community structures is not present  ($\beta=0$). In this case the effect of the media pressure on the number of clusters is remarkable only for high values of the tolerance ($\varepsilon>0.4$), when high values of the media pressure produce the increase of the number of opinion clusters.

When community structures is present ($\beta=20$), the basic scenario of
the loss of control by the strong mass media exposure, takes a different shape: 
First we can observe that the tuning of media pressure has the general role of creating, when a media pressure threshold is reached, a general bi-polarized configuration as previously described for the $\varepsilon=0.5$ case.

In Fig. \ref{fig:6}B-C we display the fraction of agents aligned
with the media opinion $\theta_M$. For both the network topologies (homophilous, \ref{fig:6}B,  and non homophilous, \ref{fig:6}C) lower values of tolerance create a stronger resistance to the dominant media independently from its pressure. At same time we observe that the maximum power of attraction  of the mass media is obtained around $pm=0.2$ and it monotonically decreases as 
the media pressure increases. However, Fig. \ref{fig:6}C shows that as a consequence of the community structures, when homophily is present, this maximum value no 
longer reaches all the population. Therefore in this case, the large mass of the counter-cluster is able to subtract support to the media.

\subsubsection*{ Central result:}
In general, independently of the network topology, the media messages are less  effective if the media pressure is too high. In the case where homophily is present in the network structure, the threshold where the media become ineffective is lower, showing that these structures are more resistant to propaganda.
Furthermore, the community structures resulting from the homophily-based networking, when facing a dominant message, aggregate the non-aligned states 
into just one second strong opinion counter-cluster, decreasing the size of dominant one. 

\section{Conclusions}

In this work three related subjects have been addressed. 
First, we showed how a network morphogenesis model, including at the same time the preferential attachment mechanism and an homophily effect, can structure networks with the same power-law degree distribution as the BA networks, but with marked communities of nodes sharing similar opinions. 

Second, the bounded confidence model has been used on such topology showing that, contrary to established ideas, homophily in social networks favors 
consensus formation: we show that the critical value of tolerance ( $\varepsilon_c= 0.5$), previously reported as the threshold for BC models to 
shift between total consensus and different opinions, loses its "universal" character when considered in more realistic networks, as the ones formed 
with community structures. In the case where homophily is present consensus is reached also for lower values of the tolerance parameter (less open-minded societies). 

Finally, the effect of mass media over the BC models with homophily scale-free networks has been reported. We showed that homophily has a double 
effect: first, it decreases the effectiveness of the media pressure, facilitating the emergence of counter opinions also for lower values of the media 
pressure. On the other hand, we showed that, when the community structures (typical of homophily-based networks) face dominant
messages, disaggregated non-aligned states converge into just one (or few) strong counter-opinion cluster, representing a
strong polarization of the opinions in the societies. Moreover, the strong polarization against the dominant message is
promoted by low values of tolerances.

Social networks and social media are nowadays the backbone of the diffusion of controversial subjects. Using data analytics tools (to personalize the advertisements) and new tools of the digital economy like the click-farms (to increase visibility to a content), new debates, often deviant from the dominant vision of the state authorities, spread everyday on the web. In several cases, when these debates, can trigger risky behaviors, the authorities answer with strong media campaigns (think for example to the case of vaccines). 

According to our findings,  these risky opinions would be naturally controlled, on long time scale. If the subject is too risky and an immediate response is needed, media campaigns are probably the worst method. Probably the best solution would be to use the same social networks to propagate the counter-messages.

A first further direction of analysis, that could be pursued in the future, concerns this last point: if traditional media apparently have lost their centrality in the communication,  how to better veicolate counter-messages to prevent risky behaviors? 
A second central direction to be addressed is the role of the click-market on the opinion formation. How the new instruments to capture users' attention are different from tradition media? Is the click economy responsible for the fact that opinions that once were considered deviant are now dominating? 

\end{document}